\documentclass[12pt]{article}
\usepackage{graphicx}
\usepackage{cite}
\usepackage{color}
\usepackage{ulem}

 \newcommand\pubnumber{}
 \newcommand\pubdate{}

\def\napoli{Department of Physics, Hanyang University,\\ 
Haengdang-Dong, Seongdong-Gu, Seoul, 133-791 Korea}

\def\Title#1{\begin{center} {\Large #1 } \end{center}}
\def\Author#1{\begin{center}{ \sc #1} \end{center}}
\def\Address#1{\begin{center}{ \it #1} \end{center}}

\newcommand\pubblock{\rightline{\begin{tabular}{l} \pubnumber\\
         \pubdate  \end{tabular}}}
\newenvironment{Abstract}{\begin{quotation}  }{\end{quotation}}
\newenvironment{Presented}{\begin{quotation} \begin{center} 
             PRESENTED AT\end{center}\bigskip 
      \begin{center}\begin{large}}{\end{large}\end{center} \end{quotation}}

\def\beq{\begin{equation}}
\def\eeq#1{\label{#1}\end{equation}}
\def\eeqn{\end{equation}}

\def\beqa{\begin{eqnarray}}
\def\eeqa#1{\label{#1}\end{eqnarray}}
\def\eeqan{\end{eqnarray}}

\let\bar=\overbar

\def\Dslash{\not{\hbox{\kern-4pt $D$}}}
\def\dslash{\not{\hbox{\kern-2pt $\del$}}}

\def\msb{{\bar{\ssstyle M \kern -1pt S}}}

 \def\pip{\mbox{$\pi^{+}$}}
 \def\pim{\mbox{$\pi^{-}$}}
 \def\piz{\mbox{$\pi^{0}$}}

 \def\kp{\mbox{$K^{+}$}}

 \def\kz{\mbox{$K^{0}$}}
 
 \def\ks{\mbox{$K^{0}_{S}$}}
\def\bz  {\mbox{$B^0$}}

\def\bp {\mbox{$B^{+}$}}

 \def\dacp{\mbox{$\Delta {\cal A}_{\rm CP}$}}
 \def\acp{\mbox{${\cal A}_{\rm CP}$}}
 \def\br{\mbox{$\cal B$}}
 \def\ab{\mbox{ab$^{-1}$}}
 \def\fb{\mbox{fb$^{-1}$}}
\begin{document}
\begin{titlepage}
 \pubblock

\vfill
\Title{Experimental status and prospects of the ``$K\pi$ puzzle''}
\vfill
\Author{Yuuji Unno}
\Address{\napoli}
\vfill
\begin{Abstract}
We present a summary of the latest measurements of branching fractions 
and direct CP violations
in the charmless hadronic two-body $B$ decays
 $B\to\kp\pim$, $\kp\piz$, $\kz\pip$ and $\kz\piz$,
performed by the Belle, BABAR, CDF and CLEO experiments. 
The interpretations and the future prospects about discrepancies, 
called ``$K\pi$ puzzle'', between theoretical expectations and measurements are discussed.
\end{Abstract}
\vfill
\begin{Presented}
The 6th International Workshop on the CKM Unitarity Triangle, 
University of Warwick, UK, 6-10 September 2010
\end{Presented}
\vfill
\end{titlepage}
\def\thefootnote{\fnsymbol{footnote}}
\setcounter{footnote}{0}
\section{Introduction}
Charmless hadronic two-body $B$ decays, $B\to K\pi$, 
provide opportunities to test the standard model(SM)
and search for direct CP violation and  
new physics beyond the SM.
$B$ mesons decay to $K\pi$ final states
through several processes listed 
in Table~\ref{tab:diagram}.
\begin{table}[htb]
\begin{center}
 \begin{tabular}{l|l} \hline \hline
  Mode         & Quark-level amplitude \\ \hline
  $\bz\to\kp\pim$  & T + P + $\rm P^C_{EW}$  \\
  $\bp\to\kp\piz$  & T + P + C + $\rm P_{EW}$ + $\rm P^C_{EW}$ + A  \\
  $\bp\to\kz\pip$  & P + $\rm P^C_{EW}$ + A \\
  $\bz\to\kz\piz$  & P + C + $\rm P_{EW}$ + $\rm P^C_{EW}$ \\\hline\hline
 \end{tabular}
\caption{Quark-level amplitudes for $B\to K\pi$. 
The contributed amplitudes are
tree(T), 
penguin(P), 
colour-suppressed tree(C),
electroweak penguin($\rm P_{EW}$),
colour-suppressed electroweak penguin($\rm P^C_{EW}$),
and weak annihilation(A).}
\label{tab:diagram}
\end{center}
\end{table}
The tree and penguin processes dominantly contribute,
and direct CP violation may arise from their interference.
The partial rate CP violating asymmetry is measured as 
$\acp = [N(\overline{B})-N(B)]/[N(\overline{B})+N(B)]$,
where $N(\overline{B})(N(B))$ is the yield
for $\overline{B}(B)$.
By ignoring subdominant contributions and/or using
isospin relations~\cite{th_gronau} one can 
predict the branching fraction({\br}),
{\acp}, and these relations among the four $B\to K\pi$ decays 
and test the SM against the measurements.
In this article we report on the {\br} and {\acp} measurements
by Belle, BABAR, CLEO and CDF, and discuss
the discrepancies from the SM, called ``$K\pi$ puzzle'', 
that have been found in the measurements,
and the future prospects.
\section{Branching fractions}
The branching fractions measured by Belle, BABAR and CLEO 
together with the averages of the three experiments
are shown in Table \ref{tab:result_br}
\cite{br_belle1,br_belle2,br_belle3,br_babar1,br_babar2,br_babar3,br_babar4,br_cleo}.
The results of three experiments are consistent with each other
and agree with theoretical calculations~\cite{th_buras} in the SM.
However, experimental results are all systematics dominated and therefore
it is crucial to reduce the systematics 
for more stringent comparison with theory.
Main systematics in Belle are due to 
charged track finding efficiency, 
kaon and pion identification, 
{\ks} reconstruction, and 
{\piz} reconstruction 
and 
similar limitations affect BABAR measurements.
\begin{table}[htb]
\begin{center}
\begin{tabular}{l|cccc}   \hline \hline
         & $\bz\to\kp\pim$              & $\bp\to\kp\piz$              & $\bp\to\kz\pip$              & $\bz\to\kz\piz$ \\ \hline
 Belle   & $19.9 \pm 0.4 \pm 0.8$       & $12.4 \pm 0.5 \pm 0.6$       & $22.8^{+0.8}_{-0.7} \pm 1.3$ & $ 8.7 \pm 0.5 \pm 0.6$ \\
 BABAR   & $19.1 \pm 0.6 \pm 0.6$       & $13.6 \pm 0.6 \pm 0.7$       & $23.9 \pm 1.1 \pm 1.0$       & $10.1 \pm 0.6 \pm 0.4$ \\
 CLEO    & $18.0^{+2.3+1.2}_{-2.1-0.9}$ & $12.9^{+2.4+1.2}_{-2.2-1.1}$ & $18.8^{+3.7+2.1}_{-3.3-1.8}$ & $12.8^{+4.0+1.7}_{-3.3-1.4}$ \\
 Average & $19.4 \pm 0.6$               & $12.9 \pm 0.6$               & $23.1 \pm 1.0$               & $9.5 \pm 0.5$ \\ \hline \hline
\end{tabular}
\caption{{Summary of \br} measurements performed by Belle, BABAR and CLEO.}
\label{tab:result_br}
\end{center}
\end{table}

\begin{figure}[htbp]
 \centering
 \includegraphics[height=5cm]{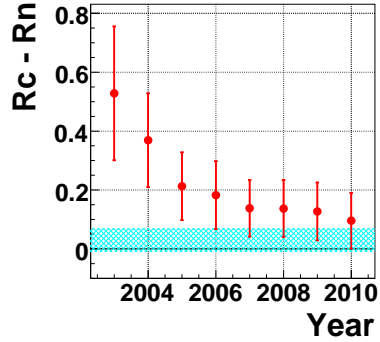}
 \caption{$R_c-R_n$ measurements as a function of year.
Blue hatched region represents $\pm 1\sigma$ uncertainty of the SM expectation.}
 \label{fig:ratio_of_br}
 \end{figure}

The ratios of {\br}, defined as 
$R_c = 2\br(\bp\to\kz\pip)/\br(\bp\to\kp\piz)$
and 
$R_n = \br(\bz\to\kp\pim)/2\br(\bz\to\kz\piz)$,
and the difference of the two ratios, $R_c-R_n$,
are sensitive to new physics.
In these ratios, there are advantages that some systematics
can be reduced both in experimental measurements and theoretical calculations.
Ref.~\cite{th_buras} predicts 
$R_c = 1.15\pm 0.03$,
$R_n = 1.12\pm 0.03$ and
$R_c-R_n = 0.03\pm 0.04$.
In 2003$-$2004, discrepancies of up to 2$\sigma$ were found 
between measurements of $R_c-R_n$ and theory predictions,
shown in Fig.~\ref{fig:ratio_of_br}.
This "puzzle" gradually disappeared once more precise measurements 
shown reduced disagreement with theory.

Current experimental results obtained 
from averages shown in Table~\ref{tab:result_br} are
$R_c = 1.12\pm 0.07$,
$R_c = 1.02\pm 0.06$ and
$R_c-R_n = 0.10\pm 0.09$, 
which are in agreement with the calculations 
in the SM at the 1$\sigma$ level.
\section{Direct CP violation}
Since both tree and penguin processes contribute
to $\bz\to\kp\pim$ and $\bp\to\kp\piz$ decays, 
sizable $\acp$ could be expected.
Moreover, $\acp$s in the $\bz\to\kp\pim$ and $\bp\to\kp\piz$ 
are expected to have approximately 
the same magnitude and sign~\cite{th_sanda}.
On the other hand, 
both $\bp\to\kz\pip$ and $\bz\to\kz\piz$ are almost pure penguin processes, 
hence no sizable asymmetries are expected in the SM.
The results of {\acp} measurements carried out by Belle, BABAR, CLEO, and CDF
with the averages are shown in Table~\ref{tab:result_acp}
\cite{acp_belle,br_belle2,br_belle3,acp_babar,br_babar2,br_babar4,acp_cdf,acp_cleo}.
\begin{table}[htb]
\begin{center}
 {\scriptsize
\begin{tabular}{l|cccc}   \hline \hline
         & $\bz\to\kp\pim$                      & $\bp\to\kp\piz$              & $\bp\to\kz\pip$              & $\bz\to\kz\piz$ \\ \hline
 Belle   & $-0.094 \pm 0.018 \pm 0.008$         & $+0.07 \pm 0.03 \pm 0.01$    & $+0.03 \pm 0.03 \pm 0.01$    & $+0.14 \pm 0.13 \pm 0.06$ \\
 BABAR   & $-0.107 \pm 0.016^{+0.006}_{-0.004}$ & $+0.030 \pm 0.039 \pm 0.010$ & $-0.029 \pm 0.039 \pm 0.010$ & $-0.13 \pm 0.13 \pm 0.03$ \\
 CDF     & $-0.086 \pm 0.023 \pm 0.009$         & $-$                          & $-$                          & $-$ \\
 CLEO    & $-0.04\ \pm 0.16 \pm 0.02$           & $-0.29 \pm 0.23 \pm 0.02$    & $+0.18 \pm 0.24 \pm 0.02$    & $-$ \\
 Average & $-0.098 \pm 0.012$                   & $+0.050 \pm 0.025$           & $+0.009\pm 0.025$            & $-0.01 \pm 0.10$ \\ \hline \hline
\end{tabular}
}
\caption{Summary of {\acp} measurements by Belle, BABAR, CLEO and CDF.}
\label{tab:result_acp}
\end{center}
\end{table}
All results are consistent across the four experiments.
$\acp(\bz\to\kp\pim)$ has reached $8\sigma$ significance,
and $\bp\to\kz\pip$ and $\bz\to\kz\piz$ show 
no significant asymmetries, given the 3\% and 10\% uncertainties, respectively. 
These follow the expectations mentioned above.
However, the measured $\acp(\bp\to\kp\piz)$ has a different
magnitude and opposite sign to $\acp(\bz\to\kp\pim)$, and
the difference is
$\Delta \acp = \acp(\bp\to\kp\piz)-\acp(\bz\to\kp\pim) = +0.148\pm 0.028$,
which 
is 5.3$\sigma$ different from zero.

Unlike in the $\bz\to\kp\pim$ decay,
in addition to the tree and penguin,
the colour-suppressed tree and
electroweak penguin processes may contribute to $\bp\to\kp\piz$
although these are expected to be much smaller than tree and penguin amplitudes.
There are several theoretical conjectures to try 
to explain this {\dacp} puzzle
with taking these contributions into account:
enhancement of the colour-suppressed tree amplitude~\cite{th_1819},
electroweak penguin contributions~\cite{th_20}, or both~\cite{th_21}.
If this effect were to be explained solely by enhancement of the
colour-suppressed tree amplitude
its amplitude would have to be larger than~\cite{th_20,th_21}
the tree amplitude. 
It means that the sizes of strong phases do not fit with factorization.
If the electroweak penguin explains the effect, 
this would indicate new physics beyond the SM~\cite{th_20,th_21,th_22}.

Further examination for new physics effect can be performed with
isospin sum rule among four {\acp}s~\cite{th_gronau2}:
 \begin{eqnarray*}
  {\acp}(\bz\to\kp\pim) + 
  {\acp}(\bp\to\kz\pip)\frac{{\cal B}(\bp\to\kz\pip)}{{\cal B}(\bz\to\kp\pim)}\frac{\tau_0}{\tau_+} = \\
  {\acp}(\bp\to\kp\piz)\frac{2{\cal B}(\bp\to\kp\piz)}{~{\cal B}(\bz\to\kp\pim)}\frac{\tau_0}{\tau_+} +
  {\acp}(\bz\to\kz\piz)\frac{2{\cal B}(\bz\to\kz\piz)}{~{\cal B}(\bz\to\kp\pim)},
 \end{eqnarray*}
 where
 $\tau_+$($\tau_0$) is a $\bp(\bz)$ meson lifetime. 
 A violation of this sum rule would be 
 an unambiguous evidence of new physics.
 With all of {\br} and {\acp} results except for {\acp}($\bz\to\kz\piz$),
 the sum rule predicts {\acp}($\bz\to\kz\piz$) to be $-0.153\pm 0.045$,
 which is consistent with 
the measurement of {\acp}($\bz\to\kz\piz$).
 \begin{figure}[htb]
 \centering
 \includegraphics[height=6cm]{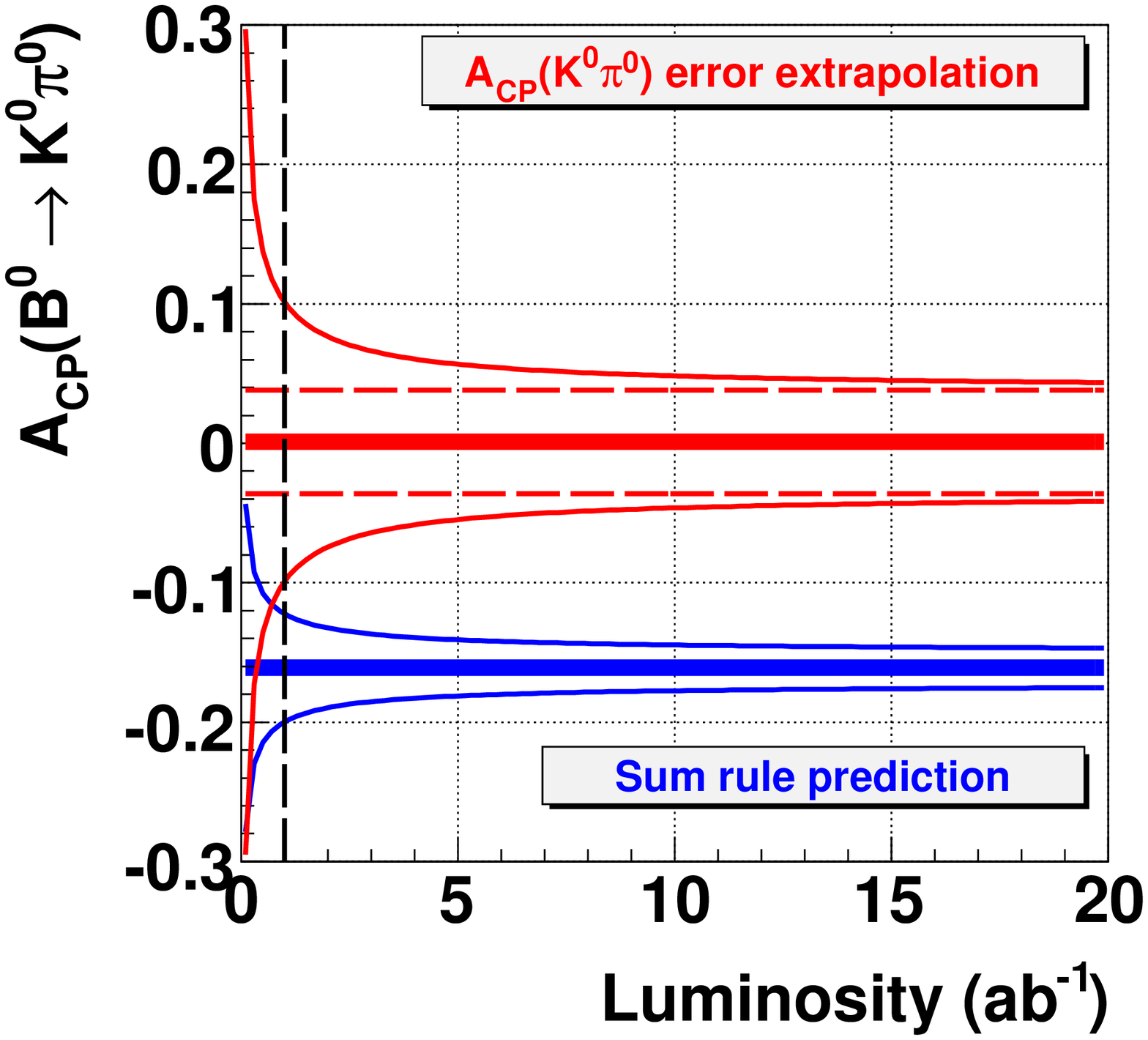}
 \caption{
Expected {\acp($\bz\to\kz\piz$) as a function of luminosity.
Red solid line, dotted line, and solid curve show central value,
systematic error, and 1$\sigma$} error including systematics, respectively,
of $\acp(\bz\to\kz\piz)$
based on current world average of 
$\acp(\bz\to\kz\piz)$.
Blue line and curve represent 
central value and 1$\sigma$ error including systematics
predicted by isospin sum rule using
current world averages except for
$\acp(\bz\to\kz\piz)$.
Dotted vertical line denotes current data set( $\sim$ 1 {\ab}).}
 \label{fig:sumrule}
 \end{figure}
More precise test of isospin sum rule will be carried out 
by two super $B$ factories, 
SuperKEKB~\cite{belle2} and Super$B$~\cite{superb}, and LHCb~\cite{lhcb}.
The super $B$ factories will improve all of four $B\to K\pi$ modes,
and although LHCb can improve $\bz\to\kp\pim$ and $\bp\to\kz\pip$ modes
the expected statistical error on the ${\acp(\bp\to\kp\pim)}$ would reach
the 0.5\% level with 1 {\fb}.
Fig.\ref{fig:sumrule} shows the expected error on the
$\acp(\bz\to\kz\piz)$
measurement and the prediction of isospin sum rule 
up to 20 {\ab} based on current world averages 
by fixing present systematics, in case of the super $B$ factories. 
At 15 {\ab}, 
the error of $\acp(\bz\to\kz\piz)$ will reach the 5\% level and 
the isospin sum rule will be violated with $3\sigma$ level
if current central value of 
$\acp(\bz\to\kz\piz)$ would remain.
However since the measurement will be systematics dominated,
an effort to reduce the systematics becomes crucial.
 %
 %
 %
 %
 %
 %
 %
 %
 %
 %
 %
 %
 %
 %
 %
 %
 %
 %
 %
 %

 %
 %
 %
 %
 %
\end{document}